# SHORT INTENSE ION PULSES FOR MATERIALS AND WARM DENSE MATTER RESEARCH

Peter A. Seidl[*], Wayne G. Greenway, Steven M. Lidia[†], Arun Persaud, Matthew Stettler, Jeffrey H. Takakuwa, William L. Waldron, and Thomas Schenkel
*Lawrence Berkeley National Laboratory, Berkeley, CA, USA*
John J. Barnard, Alex Friedman, David P. Grote
*Lawrence Livermore National Laboratory, Livermore, CA USA*
Ronald C. Davidson, Erik P. Gilson, Igor D. Kaganovich
*Princeton Plasma Physics Laboratory, Princeton, NJ USA*

---

[*] PASeidl@lbl.gov
[†] present address: Facility for Rare Isotope Beams, Michigan State University, Michigan, USA.

## Abstract

We have commenced experiments with intense short pulses of ion beams on the Neutralized Drift Compression Experiment-II at Lawrence Berkeley National Laboratory, by generating beam spots size with radius r < 1 mm within 2 ns FWHM and approximately $10^{10}$ ions/pulse. To enable the short pulse durations and mm-scale focal spot radii, the 1.2 MeV Li$^+$ ion beam is neutralized in a 1.6-meter drift compression section located after the last accelerator magnet. An 8-Tesla short focal length solenoid compresses the beam in the presence of the large volume plasma near the end of this section before the target. The scientific topics to be explored are warm dense matter, the dynamics of radiation damage in materials, and intense beam and beam-plasma physics including selected topics of relevance to the development of heavy-ion drivers for inertial fusion energy. Here we describe the accelerator commissioning and time-resolved ionoluminescence measurements of yttrium aluminium perovskite using the fully integrated accelerator and neutralized drift compression components.

## INTRODUCTION

Intense pulses of ions in the MeV range may be used to study the properties of matter from low intensity (negligible heating but collective effects due to proximate ions in time and space) up to high intensity where the target material may be heated to the few-eV range and beyond. By choosing the ion mass and kinetic energy to be near the Bragg peak, dE/dx is maximized and a thin target may be heated with a high degree of uniformity [Gr04], thus enabling high-energy density physics (HEDP) experiments in the warm dense matter (WDM) regime. The Neutralized Drift Compression Experiment (NDCX-II) was designed and built with this motivation [Fr10, Ba14].

Reproducible ion pulses (N>$10^{11}$ /bunch), with a bunch duration and spot size in the nanosecond and millimeter range meet the requirements to explore the physics topics above. The formation of the bunches generally involves an accelerator beam with high perveance and low emittance, which lends itself to studies of basic beam physics questions of general interest which are also relevant to the generation of high current and intensity beams for heavy-ion-driven inertial fusion energy production [To15, Ba13].

Furthermore, short ion pulses at high intensity (but below melting) enable pump-probe type experiments exploring the dynamics of radiation-induced defects in materials. Below, we present first results on the luminescence of scintillator crystals subject to nanosecond lithium ion pulses with peak currents ~18 A/cm$^2$. For high peak currents and short ion pulses, the response of the material to radiation may enter a non-linear regime due to the overlapping collision cascades initiated by the incident ions. These effects may be transient (no memory effect at a subsequent pulse) and short, intense pulses of ions provide an opportunity to observe the time-resolved multi-scale dynamics of radiation-induced defects [Sc13, Pe14]. In addition, by measuring the ion range during the course of the ion pulse, the effects of defects and heating on range can be observed.

Here we describe the commissioning of the NDCX-II and first target experiments with the fully integrated accelerator and drift-compression components. The first target experiments used beam pulses of Li$^+$ accelerated to 1.2 MeV and focused to r = 1 mm and duration of 2 ns FWHM. These conditions were used to characterize the ionoluminescence of yttrium aluminium perovskite (YAP).

## ACCELERATION TO 1.2 MEV AND FOCUSING

The NDCX-II accelerator is comprised of a Li$^+$ ion source, pulsed injector and induction accelerator [Wa12]. The final kinetic energy of the accelerator is presently 1.2 MeV, a four-fold increase compared to Refs. [Sc13, Li15] resulting from the integration of five high-voltage pulsers with the last acceleration modules in the linac. A significant task was to route and connect the massive, oil-filled transmission lines from the Blumleins to the induction cells.

The ion source is a filament-heated Li$^+$ aluminosilicate coating on a porous tungsten substrate heated to 1200-1250 C. The preparation of the emitter and the injector beam characteristics are described in [Se12]. The lifetime of the nearly freestanding toroidal filament has increased significantly by preparing the toroidal winding with a four-strand tungsten filament instead of a single 1.5-mm diameter strand. Operating the emitter at a slightly lower temperature (1200 C vs. 1220-1240 C) for a greater fraction of the operating time also contributed to lifetime extension. These improvements allowed operation with Li$^+$ sources in some instances for a few months (versus weeks), where the lifetime was limited by the depletion of the lithium from the source coating rather than a failure of the filament. The extracted charge from the emitter during the ~1 μs pulse was up to 30 nC. This charge is captured in a matching solenoid just before the first acceleration module. The injection energy is 135 keV.

To attain significantly greater charge/pulse and flexibility, a multicusp type plasma source is being tested and will inject ~100 nC of, for example, helium ions into the accelerator for WDM and other experiments. An advantage of helium ions is that 1.2 MeV ions will enter thin (1-2 μm) samples slightly above the Bragg peak and deposit energy uniformly through the sample, complementing WDM experiments driven by intense lasers and x-ray beams.

An induction accelerator is capable of accelerating and rapidly compressing beam pulses by adjusting the slope and amplitude of the voltage waveforms in each gap. In NDCX-II, this is accomplished with 12 compression and acceleration waveforms driven with peak voltages ranging from 15 kV to 200 kV and durations of 0.07-1 μs (Fig. 1).

The first seven acceleration cells are driven by custom spark-gap, switched, lumped element circuits tuned to produce the required cell voltage waveforms. These waveforms ("compression" waveforms because of the characteristic triangular shape) have peak voltages ranging from 20 kV to 50 kV. An essential design objective of the compression pulsers is to compress the bunch to <70 ns so that it can be further accelerated and bunched by the 200-kV Blumlein pulsers which drive the last five acceleration cells. The locations of compression, Blumlein and diagnostics in the 27-cell accelerator and target station are shown in Table 1.

Each lattice period has a pulsed solenoid with a maximum operating field of 2.5 Tesla and a beam tube diameter of 4 cm. The transverse dynamics are space-charge-dominated; thus to maintain an approximately constant beam radius through the accelerator, the focusing fields are increased in the latter stages of the accelerator to balance the greater space charge forces. Pulsed magnetic dipoles are co-located with seven solenoids through the lattice, producing several milliradian kicks to the beam centroid.

The first three Blumleins generate ≤200 kV flat waveforms across the acceleration gaps. The final two are ramped waveforms, generating the final velocity ramp and energy gain to the beam before the drift compression section. Each Blumlein pulser is charged with a thyratron-switched charging chassis.

These five Blumlein-driven cells increase the beam energy from approximately 0.3 MeV to 1.2 MeV in a few meters. At the exit of the final acceleration gap the bunch duration is 30-40 ns. Thus the physical bunch length is decreased from the injector ($l_{bunch} = \tau_{inj} \cdot v_{beam} = 1.4 m$) by nearly an order of magnitude to 0.17 m at end of the accelerator.

| Cell# | Function | Z (cm) |
|---|---|---|
| 1 | Compression 1 | 72 |
| 2 | Compression 2 | 100 |
| 3 | BPM 1 | 128 |
| 4 | Current monitor 1 | 155 |
| 5 | Current monitor 2 | 183 |
| 6 | Compression 3 | 211 |
| 7 | BPM 2 | 239 |
| 8 | Current monitor 3 | 267 |
| 9 | Current monitor 4 | 295 |
| 10 | Compression 4 | 323 |
| 11 | BPM 3 | 351 |
| 12 | Current monitor 5 | 379 |
| 14 | Current monitor 6 | 435 |
| 15 | Compression 5 | 463 |
| 16 | BPM 4 | 491 |
| 17 | Compression 6 | 519 |
| 18 | Current monitor 7 | 547 |
| 19 | Compression 7 | 575 |
| 20 | BPM 5 | 603 |
| 21 | Current monitor 8 | 630 |
| 22 | Blumlein 1 (flat) | 658 |
| 23 | Blumlein 2 (flat) | 686 |
| 24 | BPM 6 | 714 |
| 25 | Blumlein 3 (flat) | 742 |
| 26 | Blumlein 4 (ramped) | 770 |
| 27 | Blumlein 5 (ramped) | 798 |
| | BPM 7 | 826 |
| Target | Beam-target experiments | 989 |

Table 1: The compression and Blumlein cells are interspersed with diagnostic stations in the 27-cell accelerator lattice. Each lattice period contains a focusing solenoid. The location (Z) is with respect to the ion-source-emitting surface.

To achieve repeatable ns-bunches on target, the timing of the induction cells, and in particular the Blumleins must be controlled to <5 ns. The effect of timing jitter on the beam quality was simulated with ensembles of 3D particle in cell simulations with measured waveforms [Gr14]. Figure 2 shows a pressure scan of a Blumlein spark gap, which was stabilized to have 2.4 ns jitter.

In the final drift section, the bunch has a head-to-tail velocity ramp that further compresses the beam by an order of magnitude. The space-charge forces are sufficiently high at this stage that externally generated plasma is needed to neutralize the beam self-field and to enable focusing and bunching of the beam to the millimeter and nanosecond range, limited mainly by emittance and chromatic aberrations.

Most of the flight path through the neutralized drift compression section is through plasma generated by a cylindrical ferro-electric plasma discharge [Gi13]. The electron charge distribution from the overdense plasma rearranges to shield the beam potential. Thus, the coasting beam compresses with negligible space-charge repulsion. The plasma sources, final focusing solenoid and the target chamber are shown in Fig. 3.

## BEAM DIAGNOSTICS

The lattice period is short with high occupancy: the effective field length of the solenoid focusing magnets is 19 cm in a 28-cm lattice period. Except at the end of the accelerator, non-intercepting diagnostics are used instead of intercepting slit scanners and Faraday cups to monitor current, centroid and particle loss.

Capacitive-coupled beam position monitors (BPM) detect the beam centroid at seven locations in the accelerator. The inner diameter of the BPMs is matched to the 7.9-cm accelerator beam tube, and the length is 1.27 cm, giving good time resolution to monitor the variation of the centroid position throughout each pulse. The BPM7 at the exit of cell #27 is longer (8.1 cm) which provides enhanced signal to noise at the expense of decreased time resolution. The $Li^+$ transit time through the BPM is 14 ns, which is acceptable for monitoring the centroid of the 20-40 ns bunches. BPM-7 is upstream of the ferroelectric plasma so biasing the electrodes enables suppression of back-streaming electrons from the plasma source into the accelerator. The BPMs, each segmented into four independent quadrants, are used to monitor the first moment of the beam and determine the dipole fields. This is illustrated in Fig. 4, where tuning the dipoles decreases the first moment offsets from 5-10 mm to <3 mm. The centroid history of the beam pulse at each station is recorded – in some cases the locus of points has a nearly closed pattern resembling the "corkscrew" pattern caused by the smooth variation of the betatron period due to the applied velocity ramp of the compressing beam [Ch92].

Inductive current monitors (CM) at eight other lattice positions monitor the beam current. The ion beam current is the primary of a 1:1 transformer (using inactive induction cell cores) and the voltage across the secondary winding is recorded. Figure 5 shows beam current waveforms from the CMs and BPMs where fivefold current amplification due to the acceleration and compression of the beam bunch is evident.

The target is housed in a dedicated diagnostic station. Stepper motors position the targets to enable an array of successive beam pulses on a single sample without opening the vacuum system. Before these beam-target shots, a fast Faraday cup measures the beam current [Se06]. The ion transit time from suppressor electrode to the collector is low and the capacitance are low, thus <1ns resolution is expected. Aligned hole plates (2% transparent) and a metal enclosure exclude the neutralizing plasma and electrical noise.

The transverse distribution of the beam is detected with a 30-mm alumina ($Al_2O_3$) scintillator with a gated CCD camera, providing a pixel resolution on the object plane of 0.2 mm. Excellent signal to noise requires gating the camera narrowly around the beam, and suppressing the contribution of background light from the nearby plasma sources and from the ion source.

A streak-spectrometer characterizes ion-induced optical emission of target samples [Ni13]. This will serve as a temperature diagnostic in the upcoming WDM experiments, with a temporal resolution significantly shorter than the ~1 ns beam pulses. It has a spectral span of ≈400 nm, and the grating and streak cathode have a sensitivity cutoff near $\lambda=200$ nm. Light is collected downstream in a moveable motorized glass lens system in the vacuum system and exits the target chamber via a multi-mode optical fiber aimed at the grating spectrometer entrance. The radiation environment surrounding the target chamber is benign, and sensitive diagnostics can be placed in close proximity to the target. This is because the total energy in the ion pulse is ≤0.1 J, and ions are generated and accelerated without generating large electromagnetic pulses compared to for example, ion pulses from high-power laser-plasma acceleration [Sn00].

## BEAM-TARGET EXPERIMENTS

Reproducibility is a great advantage for many beam target experiments. It allows accurate tuning of the beam parameters to establish the desired fluence at the target before irradiating target samples. Repeated shots of the beam characterized the stability of the ion source beam distribution, the correction dipoles and 28 focusing solenoids. Using the scintillator and gated CCD camera at the end of the last accelerator cell to detect a coasting ~0.14 MeV beam, the beam centroid varied by $\sigma_{x,y} <$ 0.2 mm over 50 shots. This is insensitive to beam current variations of ±5%, which we attribute to variations in the emitter temperature.

Centroid variations at the target plane may be caused by any significant fluctuations from the plasma sources and from the final focusing magnet strength. However, the centroid variation was independently measured to be

$\sigma_{x,y} < 0.1$ mm, consistent with the upstream measurements and within the pixel resolution of the CCD camera optics. The variation of the peak light emission amplitude was $\sigma_A/A < 7\%$. This level of stability enables tuning the beam so that single-shot experiments will be accurately located on the target samples.

Figure 6 shows the beam current distribution at the target plane. A sharp 0.6-Ampere peak demonstrates that the beam pulse compressed 5-10x in the final 1.6-meter neutralized drift section. The current waveforms, measured before and after iterative scanning to improve the timing of the acceleration pulsers, show the importance of controlling the compression and Blumlein pulsers to a few nanoseconds, thereby achieving 1.9 ns FWHM. The integrated charge in these pulses is 1.1-1.4 nC.

The transverse distribution of the beam is sharply focused by the 0.2-m focal length final focusing solenoid, with the best focus achieved using a peak field of $7.7 < B < 8.3$ Tesla (Fig. 7). The width of the lineout is 1.3 mm FWHM. At the peak, the dose rate is $1 \times 10^{20}$ ions/cm$^2$/s, with ion fluence per shot of $2 \times 10^{11}$ ions/cm$^2$.

*Ionoluminescence of YAP scintillator*

We have measured the ionoluminescence of cerium doped yttrium aluminium perovskite, (YAP:Ce, YAlO$_3$). YAP:Ce is a fast scintillator with a 25 ns fluorescence decay time with a luminosity of about $10^4$ photons/MeV. Its luminescence properties are mostly characterized in the literature with lasers or gamma radiation [Mo97, Ha05, Ba91, Le95, De00]. For the 1.2 MeV Li$^+$ ions in this experiment, electronic stopping dominates over nuclear stopping power with a ratio of about 50:1.

The light emission was recorded with the fiber-optic coupled streak-spectrometer triggered just prior to the arrival of the 1-2 ns beam at the YAP scintillator target. The front of the target was coated with a 50 nm thick film of aluminium to increases collection of emitted light into the collection optics. The YAP crystal thickness was 1 mm. The time resolution of the streak camera was set to 2 ns with a range of 100 ns. A single-shot measurement in Fig. 8 shows time-dependent light emission as a function of wavelength. The structure in the spectrum is better resolved by summing over several beam shots, as shown in Fig. 9. The peak near 400 nm is attributed to a convolution of the broad emission peak due to the d→f transition in Ce$^{3+}$ [Ba91] with the wavelength cut-off of the optical fiber at 400 nm. Separately, analysing the emission near 400 and 440 nm we find exponential decay constants of 30 and 24 ns, respectively, similar to photon luminescence results for YAP [Ba91, Le95, De00]. This structure will be explored in future experiments in relation to earlier observations of excitation specific emission structures from luminescent materials [To13]. This result illustrates the concept and potential of time-resolved single-shot measurements, where the time resolution is limited by the diagnostics and the availability of intense, short, ion pulses enables studies of multi-scale materials response to radiation.

## CONCLUSION

All components of the Neutralized Drift Compression Experiment-II (NDCX-II) have been integrated and commissioned. The first results demonstrate the generation of intense nanosecond ion pulses with reproducible peak dose rates of $10^{20}$ ions/cm$^2$/s. Higher dose rates are anticipated with a new plasma ion source, which will enable target heating to the warm dense matter regime with protons and helium ions.

We note that laser-plasma generated protons and ions are capable of generating very high dose rates, and thus generate WDM conditions [Pa03]. Target normal sheath acceleration and other laser plasma acceleration mechanisms are creating opportunities for other applications [Ro14]. The NDCX-II facility is complementary, offering a relatively high repetition rate, excellent shot-to-shot reproducibility, and a low-noise environment near the target plane.

The ionoluminescence of YAP shows an exponential decay (24 and 30 ns, depending on the wavelength) of light emission, similar to values in other experiments but with spectral structure that we do not find in other published work on the topic. This demonstrates the novel opportunity to explore time-resolved, multi-scale dynamics of radiation-induced defects using the NDCX-II facility.

## ACKNOWLEDGMENT

We are grateful to Thomas Lipton, Chip Kozy, Takeshi Katayanagi and Ahmet Pekedis for their technical support. Their efforts on the recent enhancements to NDCX-II included the final installation of the Blumlein pulsed power system which quadrupled the ion kinetic energy. The neutralized drift compression section, target chamber and diagnostics enabled the generation of nanosecond and millimeter pulses. This work was supported by the Director, Office of Science, Office of Fusion Energy Sciences, of the U.S. Department of Energy under Contract No. DE-AC02-05CH11231.

# FIGURES

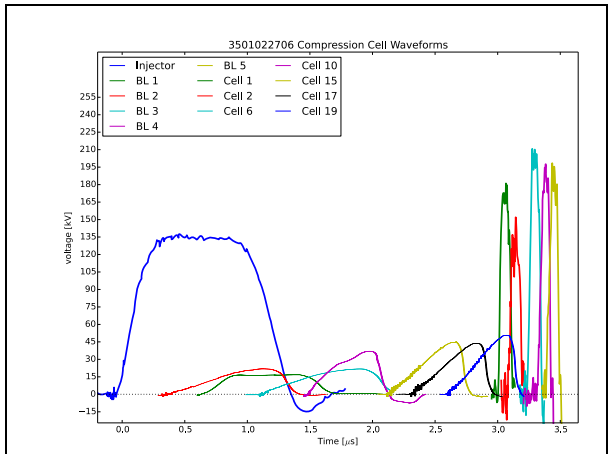

Figure 1: Voltage waveforms for each of the 12 acceleration gaps in NDCX-II.

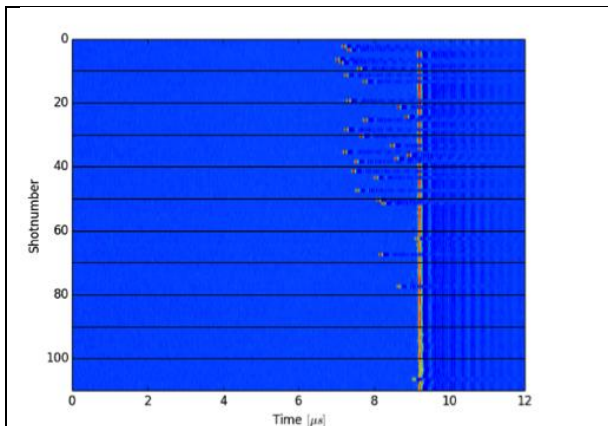

Figure 2: Blumlein timing for 110 successive shots. The voltage for shot numbers 1-80 was increased until the timing stabilized to 2.4 ns with only one outlier in 30 shots.

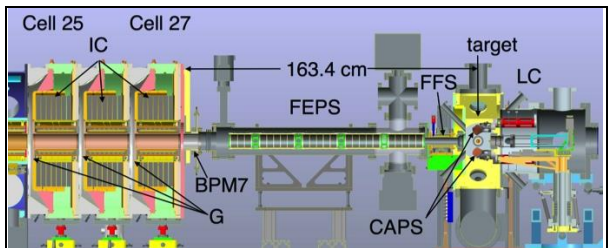

Figure 3: Side view of the end of the accelerator (cells 25-27) followed by the drift compression section, final focusing and target chamber. FEPS: Ferroelectric plasma source; G: acceleration gaps; LC: light collection and fiber optics diagnostics.

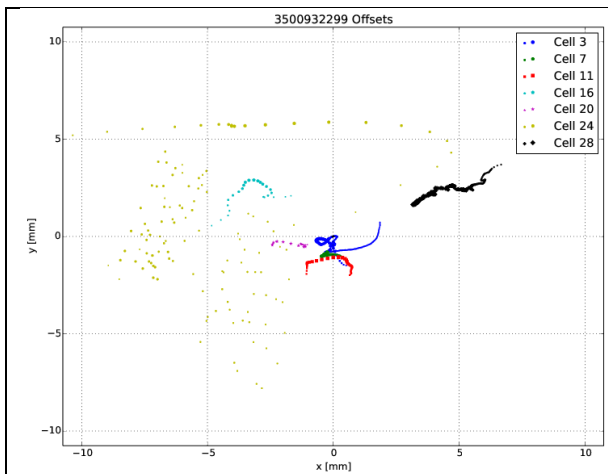

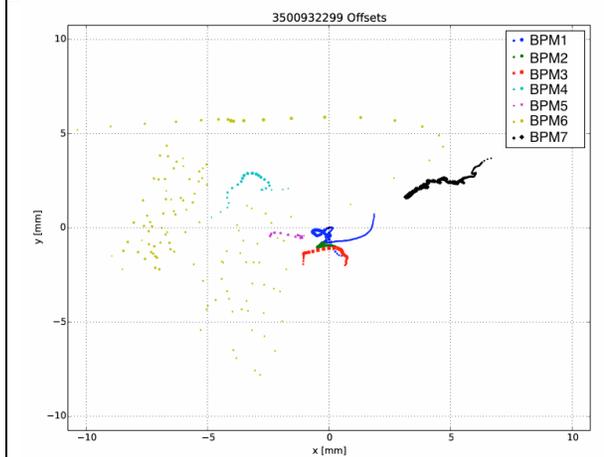

Figure 4: The centroid motion for a single bunch is shown at each of seven BPMs before (top) and after (bottom) dipole field adjustments.

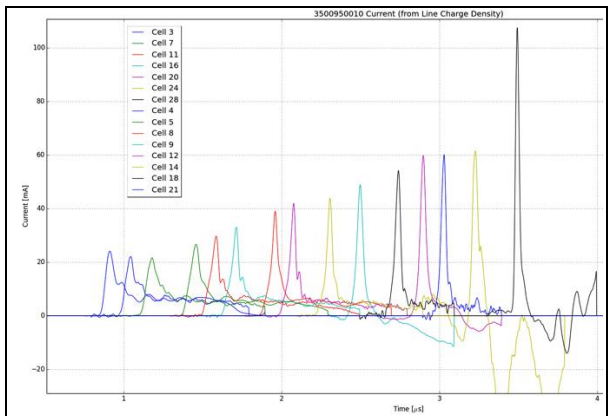

Figure 5: Current waveforms along the accelerator derived from the 15 capacitive beam position monitors and the inductive current monitors.

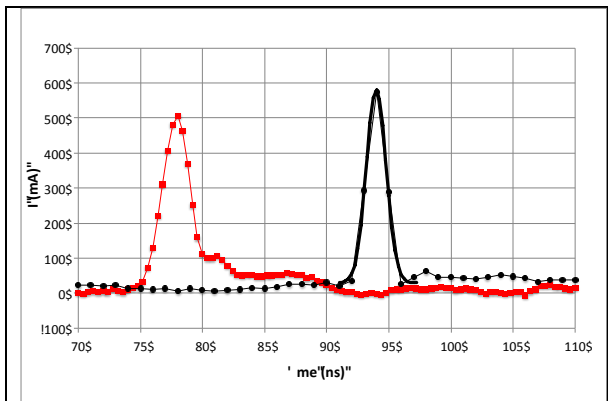

Figure 6: Faraday cup signals for two pulses. The pulse on the right followed induction cell tuning, which decreased the shoulder after the main peak. The fitted FWHM are 2.7 ns and 1.9 ns.

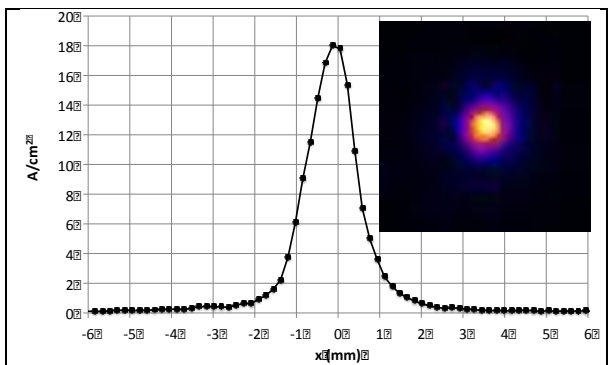

Figure 7: Beam distribution at the target plane measured with the Al2O3:Cr scintillator and gated CCD camera. The curve is a horizontal lineout near the center of the beam and the inset image (10x10 mm$^2$) shows the distribution details. (CCD pixel size = 0.17 mm on the scintillator).

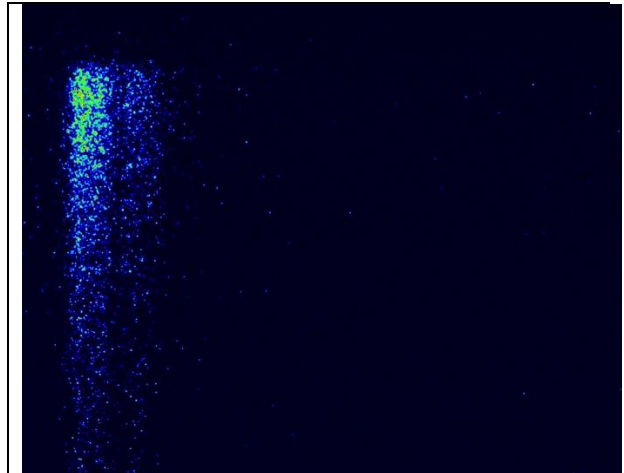

Figure 8: Ionoluminescence of YAP:Ce measured with a streaked optical spectrometer. The vertical time range is 100 ns and the abscissa range is $363 < \lambda < 800$ nm. The optical fiber sensitivity drops sharply for $\lambda < 400$ nm.

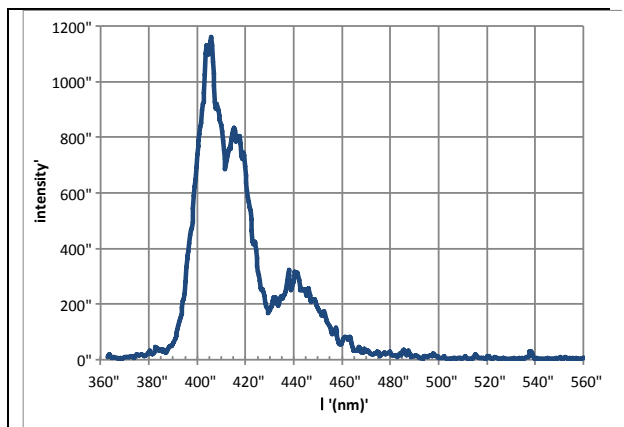

Figure 9: Ionoluminescence spectrum from YAP:Ce due to 1.2 MeV Li$^+$ ions bombarding in a few ns pulse. The diagnostic sensitivity decreases rapidly for $\lambda < 400$ nm due to the fiber optic coupling from the target chamber to the streak camera.